# *tinyVAST*: R package with an expressive interface to specify lagged and simultaneous effects in multivariate spatio-temporal models

James T. Thorson[1,*], Sean C. Anderson[2], Pamela Goddard[3], Christopher N. Rooper[2]

[1] Resource Ecology and Fisheries Management, Alaska Fisheries Science Center

[2] Pacific Biological Station, Fisheries and Oceans Canada, Nanaimo, British Columbia V9T 6N7, Canada

[3] Lynker Technologies, LLC, Under contract to Alaska Fisheries Science Center, National Marine Fisheries Service, National Oceanic and Atmospheric Administration, 7600 Sand Point Way, Seattle, WA 98115, USA

* James.Thorson@noaa.gov

Running header: expressive multivariate spatio-temporal model

**Abstract**

Multivariate spatio-temporal models are widely applicable, but specifying their structure is complicated and may inhibit wider use. We introduce the R package *tinyVAST* from two viewpoints: the software user and the statistician. From the user viewpoint, *tinyVAST* adapts a widely used formula interface to specify generalized additive models, and combines this with arguments to specify spatial and spatio-temporal interactions among variables. These interactions are specified using arrow notation (from structural equation models), or an extended arrow-and-lag notation that allows simultaneous, lagged, and recursive dependencies among variables over time. The user also specifies a spatial domain for areal (gridded), continuous (point-count), or stream-network data. From the statistician viewpoint, *tinyVAST* constructs sparse precision matrices representing multivariate spatio-temporal variation, and parameters are estimated by specifying a generalized linear mixed model (GLMM). This expressive interface encompasses vector autoregressive, empirical orthogonal functions, spatial factor analysis, and ARIMA models. To demonstrate, we fit to data from two survey platforms sampling corals, sponges, rockfishes, and flatfishes in the Gulf of Alaska and Aleutian Islands. We then compare eight alternative model structures using different assumptions about habitat drivers and survey detectability. Model selection suggests that towed-camera and bottom trawl gears have spatial variation in detectability but sample the same underlying density of flatfishes and rockfishes, and that rockfishes are positively associated with sponges while flatfishes are negatively associated with corals. We conclude that *tinyVAST* can be used to test complicated dependencies representing alternative structural assumptions for research and real-world policy evaluation.

**Introduction**

Multivariate spatio-temporal models are applicable to many questions throughout ecology, human health, econometrics, and geosciences (Thorson and Kristensen 2024). For example, these models are used to study:

1. *Community assembly* when applying joint species distribution models to estimate how species traits and local environmental conditions explain variation to density for multiple species across space and time (Ovaskainen et al. 2017);
2. *Species and/or temporal ordination* when identifying a reduced set of variables that can collectively explain biophysical variation for many processes across space and time. When simplifying dynamics over time, estimated indices (and spatial responses) are often called "empirical orthogonal functions" (Wikle and Cressie 1999), although ordination can instead (or simultaneously) be applied to identify a small number of "species archetypes," where density for each species is estimated as a weighted average of different archetypes (Latimer et al. 2009);
3. *Species interactions*, i.e., estimating how species density for one species affects local productivity for other species. Species interactions can be approximated by fitting a vector-autoregressive model to log-densities (Wootton and Emmerson 2005), and the estimated "community matrix" then represents many important properties about community stability (Ives et al. 2003).

In the following, we will use the term "multivariate spatio-temporal models" for this large family of analyses.

In the following, we introduce a new package *tinyVAST* for the R statistical environment (R Core Team 2023) that provides a simple and user-friendly interface for multivariate spatio-temporal models, and which builds upon a wide range of existing models and software (Table 1). We explain the model from two contrasting viewpoints: the software user (to view the user-interface) and the statistician (to view the statistical machinery). We then introduce a case-study example involving data from two sampling programs for seafloor-associated fishes as well as biogenic habitat (sponges and corals). We outline eight alternative hypotheses regarding how these sampling programs might relate to one another, fit all models using *tinyVAST*, and use conventional model-selection to identify a parsimonious description of habitat associations for these species.

**Viewpoint #1: User-interface and feature set**

We aim to develop software that involves a minimal set of user-level inputs, but which can still encompass a wide range of spatio-temporal model configurations. We specifically want the user-interface that resembles existing regression software in R, e.g., *glmmTMB* for hierarchical models (Brooks et al. 2017) and *mgcv* for generalized additive models (Wood 2017). Similar to these well-known regression packages, *tinyVAST* requires a data-frame `data` that contains any combination of continuous and categorical variables, and then an argument `formula` that identifies the response variables, some transformation of predictor variables, and an offset variable. This regression interface allows new users to quickly fit a model and then interrogate model components using some combination of S3-generic functions, e.g., `predict`, `summary`, etc.

By itself, however, the well-known `formula` interface for regression packages is not sufficiently expressive, e.g., to represent dependencies among predictor variables that are relevant during mediation, instrumental variable, or attribution analysis (Pearl 2009). We therefore predict the mean of a response variable as the sum of three different terms:

1. *Generalized additive model*: The user specifies a formula for a generalized additive model (GAM) that is interpreted by R-package *mgcv*. If other model components are missing, *tinyVAST* estimates parameters that are similar to *mgcv*, with small differences resulting from different methods for parameter estimation;
2. *Spatial interactions among variables*: The user can specify interactions among variables at a given site (or for spatially correlated latent variables) using "arrow" notation derived from path analysis (Wright 1934), based on the interface from R-package *sem* (Fox et al. 2020);
3. *Spatio-temporal interactions among variables*: The user can specify simultaneous and lagged interactions among variables over time using an expanded "arrow-and-lag" notation that is derived from R-package *dsem* (Thorson et al. In press).

Combining these three terms results in a flexible and generic ("expressive") model structure, and each term is useful individually (or in combination) depending upon study goals.

*Formula interface for exogenous covariates*

To specify the GAM component of *tinyVAST*, we use the formula interface from package *mgcv*. This interface is appropriate to estimate how exogenous variables affect the specified response, and allows the user to select from a wide range of nonlinear response functions. The *mgcv* formula includes univariate splines, multivariate splines, random intercept and random-slope models, as well as standard functionality for polynomial basis expansion of fixed effects. Most common options are supported within *tinyVAST*, and these options are well described elsewhere (Zuur et al. 2009, Wood 2017).

*Arrow notation for interactions among variables*

Specifying spatial interactions among variables allow users to estimate one or more spatial variables that are constant over time, e.g., representing the long-term utilization distribution that results from species interactions. To represent interactions among variables over time, we draw upon the existing notation for structural equation models and path analysis (Wright 1934). In its simplest form, this "arrow notation" specifies a set of linear dependencies among variables and is written using multiple lines of text, where each line specifies a dependency (i.e., coefficient linking two variables, or exogenous variance). Each line then includes three arguments separated by commas. The first argument specifies which variables are involved, where a one-headed arrow indicates a slope parameter and a two-headed arrow indicates a variance or covariance parameter. The second argument then defines a name for the parameter, and the (optional) third defines a starting value. This interface allows users to define multiple parameters with the same value (by using the same parameter name in multiple lines), or fix parameters at a hypothesized value a priori (by naming the parameter NA, and fixing it at the starting value).

For example, an analyst might envision a linear model with independent variable $X$ and dependent variable $Y$, where variation in $X$ causes variation in $Y$. This is specified by defining a slope parameter $\beta_1$ measuring how a change in $X$ causes a change in $Y$ (using a one-headed arrow in first line of Eq. 1), and then two variance parameters $\sigma_X$ and $\sigma_Y$ (using two-headed arrows in second and third lines):

$$X \rightarrow Y, \text{beta\_1}, 1$$
$$X \leftrightarrow X, \text{sigma\_X}, 1 \tag{1}$$

$$Y \leftrightarrow Y, \text{sigma\_Y}, 1$$

This arrow notation specifies a linear model:

$$X \sim \text{Normal}(\mu_X, \sigma_X^2)$$
$$Y \sim \text{Normal}(\mu_Y + \beta_1 X, \sigma_Y^2). \tag{2}$$

However, arrow notation can also be used to specify complex and recursive dependencies among variables. For example:

$$X \to X, \text{beta\_1}$$
$$X \to Y, \text{beta\_1} \tag{3}$$
$$Y \to Z, \text{beta\_2}$$

specifies an impact of $X$ on $Y$, and an impact of $Y$ on $Z$, such that $X$ has an indirect impact on $Z$ with magnitude $\beta_1 \beta_2$. However, it also specifies a direct impact of $X$ on $Z$, such that the total impact of $X$ on $Z$ is $\beta_1 \beta_2 + \beta_3$. By default, the software also includes a variance parameter for each variable (in this case, $\sigma_X$, $\sigma_Y$, and $\sigma_Z$), such that it results in a set of linear models:

$$X \sim \text{Normal}(\mu_X, \sigma_X^2)$$
$$Y \sim \text{Normal}(\mu_Y + \beta_1 X, \sigma_Y^2) \tag{2}$$
$$Z \sim \text{Normal}(\mu_Z + \beta_2 Y + \beta_3 X, \sigma_Z^2).$$

Importantly, this model can give useful insight about how an exogenous change in mediator Y is expected to result in a change in response Z. We use this "arrow" notation to specify interactions among variables that are constant over time.

*Arrow-and-lag notation for interactions among variables over time*

Specifying spatio-temporal interactions among variables over time allows *tinyVAST* to estimate short-term deviations from long-term average conditions. However, arrow notation does not include any concept of time, and cannot distinguish between simultaneous versus lagged effects. We therefore follow Thorson et al. (In press) in defining an expanded "arrow-and-lag" notation to allow users to specify interactions among variables over time using multiple lines of text. This text follows the same format as the "arrow notation," except it adds an additional (second) argument representing the lag involved, and now involves three or four arguments per line.

For example, the analyst might approximate species interactions by estimating a lagged impact of each species on per-capita productivity for every other species (Ives et al. 2003). This requires specifying a bivariate vector autoregressive (a.k.a., cross-lagged) model:

$$
\begin{aligned}
&X \rightarrow X, 1, \text{beta\_xx}, 0.1 \\
&X \rightarrow Y, 1, \text{beta\_xy}, 0.1 \\
&Y \rightarrow X, 1, \text{beta\_yx}, 0.1 \\
&Y \rightarrow Y, 1, \text{beta\_yy}, 0.1 \, .
\end{aligned}
\tag{4}
$$

where this then specifies:

$$
\begin{aligned}
X_t &\sim \text{Normal}(\mu_X + \beta_{xx} X_{t-1} + \beta_{yy} Y_{t-1}, \sigma_X^2) \\
Y_t &\sim \text{Normal}(\mu_Y + \beta_{xy} X_{t-1} + \beta_{yy} Y_{t-1}, \sigma_Y^2),
\end{aligned}
\tag{5}
$$

where this involves estimating four cross-lagged interactions, and can instead be written as $(X_t, Y_t)^T = \mathbf{B}(X_{t-1}, Y_{t-1})^T + \boldsymbol{\epsilon}_t$ where $\boldsymbol{\epsilon}_t \sim \text{MVN}(\mathbf{0}, \text{diag}(\sigma_X^2, \sigma_Y^2))$. This notation then allows the user to specify (among other models) a vector-autoregressive (VAR), autoregressive integrated moving average (ARIMA), dynamic factor analysis (DFA), and difference-in-differences model (Table 2).

*Spatial objects*

The user can specify an S3 object `spatial_graph` that is used to define a spatial precision matrix $\mathbf{Q}_{spatial}$, where the matrix inverse $\mathbf{Q}_{spatial}^{-1}$ represents the covariance among different sites. Options include the following:

1. *Continuous spatial domain*: Using the R-package *fmesher* (Lindgren 2023), the user can apply the stochastic partial differential equation (SPDE) approximation to a Matérn correlation function over a two-dimensional spatial domain (Lindgren et al. 2011). Package *tinyVAST* uses *fmesher* to construct the projection matrices that apply bilinear interpolation to calculate random variables at any sampled or predicted location within the defined spatial domain.

2. *Areal spatial domain*: Using the R-package *igraph* (Csardi and Nepusz 2006), the user can instead construct a graph representing polygons or other areal data. Package *tinyVAST* constructs the adjacency matrix and applies a simultaneous autoregressive (SAR) model to construct the spatial precision matrix (Ver Hoef et al. 2018). The user can specify an unconnected graph, which then eliminates any spatial correlation among modeled locations, resulting in a multivariate time-series model.

3. *Stream network domain*: Using the R-package *sfnetwork* (van der Meer et al. 2023), the user can construct a graph representing locations along a stream network (Hoef et al. 2006). Package *tinyVAST* then constructs the precision matrix representing an Ornstein-Uhlenbeck process (Charsley et al. 2023).

If the user does not specify the argument `spatial_graph`, then *tinyVAST* assumes that all observations are from a single site, and therefore reverts to a univariate time-series model (see summary of arguments in Table 3). Alternatively, the user can specify that all sites are independent using the package *igraph*, and this then reverts to a multivariate time-series model. A fitted model can then be explored using standard S3-generic functions (Table 4).

**Viewpoint #2: Statistical model structure**

Package *tinyVAST* fits a generalized linear mixed model (GLMM) where random effects are specified using Gaussian Markov random fields (GMRFs). This allows

Given the arrow notation representing interactions among $C$ variables, *tinyVAST* constructs a sparse precision matrix $\mathbf{Q}_{sem}$ with dimension $C \times C$:

$$\mathbf{Q}_{sem} = (\mathbf{I} - \mathbf{P})^t \mathbf{\Gamma}^{-t} \mathbf{\Gamma}^{-1} (\mathbf{I} - \mathbf{P}), \tag{6}$$

where $\mathbf{P}$ is the sparse $C \times C$ matrix of path coefficients (specified using one-headed arrows in the arrow notation) and $\mathbf{\Gamma}$ is the sparse $C \times C$ matrix of exogenous covariance parameters (two-headed arrows in arrow notation). Package *tinyVAST* then constructs a separable precision for a $S \times C$ matrix $\mathbf{\Omega}$ of $S$ sites and $C$ variables:

$$\text{vec}(\mathbf{\Omega}) = \text{MVN}\big(\mathbf{0}, \mathbf{Q}_{sem}^{-1} \otimes \mathbf{Q}_{spatial}^{-1}\big). \tag{7}$$

Similarly, the arrow-and-lag notation representing interactions among variables over time is used to construct a sparse $CT \times CT$ precision matrix $\mathbf{Q}_{dsem}$. This again is used to specify a separable precision for a $S \times C \times T$ three-dimensional array $\mathbf{E}$ across $S$ sites, $C$ variables, and $T$ times:

$$\text{vec}(\mathbf{E}) = \text{MVN}(\mathbf{0}, \mathbf{Q}_{dsem}^{-1} \otimes \mathbf{Q}_{spatial}^{-1}). \tag{8}$$

Finally, the formula interface is used to construct an $I \times J$ design matrix $\mathbf{X}$ for fixed effects, and an $I \times K$ design matrix $\mathbf{Z}$ for random effects, where random effects represent spline coefficients or other shrinkage priors.

However, some specifications using "arrow notation" for the space-variable interaction $\mathbf{\Omega}$ or "arrow-and-lag notation" the space-variable-time interaction $\mathbf{E}$ will result in $\mathbf{Q}_{sem}$ and/or $\mathbf{Q}_{dsem}$ not being full rank (see Table 2 for examples). In these cases, the covariance $\mathbf{Q}_{sem}^{-1}$ and/or $\mathbf{Q}_{dsem}^{-1}$ does not exist. The user can instead specify an alternative parameterization:

$$\text{vec}(\widetilde{\mathbf{\Omega}}) = \text{MVN}(\mathbf{0}, \mathbf{I}_C \otimes \mathbf{Q}_{spatial}^{-1})$$
$$\mathbf{\Omega} = (\mathbf{I} - \mathbf{P})^{-t} \mathbf{\Gamma}^t \widetilde{\mathbf{\Omega}}, \tag{9}$$

where $\mathbf{I}_C$ is a $C \times C$ identity matrix such that $\mathbf{I}_C \otimes \mathbf{Q}_{spatial}^{-1}$ has full rank (and a similar reparameterization is done for $\mathbf{E}$). This "projection" parameterization ensures that coefficients $\widetilde{\mathbf{\Omega}}$ and $\widetilde{\mathbf{E}}$ are full rank, while their projected values $\mathbf{\Omega}$ and $\mathbf{E}$ have the reduced-rank structure specified by the user.

The expected value $\mu_i$ for sample $i$ is then calculated via by applying an inverse-link function $g_i^{-1}(.)$ to a linear predictor, which is calculated in turn by assembling all three components:

$$g_i(\mu_i) = \underbrace{\mathbf{X}_i \boldsymbol{\alpha} + \mathbf{Z}_i \boldsymbol{\gamma}}_{GAM} + \underbrace{\mathbf{A}_i \mathbf{\Omega}_{c_i}}_{SEM} + \underbrace{\mathbf{A}_i \mathbf{E}_{c_i, t_i}}_{DSEM}, \tag{10}$$

where $\mathbf{A}$ is a sparse $I \times S$ matrix that projects spatial variables from $S$ modeled locations to the $I$ sampled locations, $\mathbf{\Omega}_{c_i}$ is the vector across $S$ locations from $\mathbf{\Omega}$ for the variable $c_i$ measured in sample $i$, and $\mathbf{E}_{c_i, t_i}$ is the vector across $S$ locations from $\mathbf{E}$ for the variable $c_i$ and time $t_i$. We allow the user to specify multiple link-functions and/or distributions associated with different partitions of the data (i.e., different variables). Specifically, $g_i(.)$ is the link function specified for sample $i$, and a probability density function $f_i(.)$ is also specified for each sample:

$$Y_i \sim f_i(\mu_i, \theta_i), \tag{11}$$

where $\theta_i$ is the dispersion parameters associated with sample $i$. Finally, blocks of random effects also follow a specified distribution, which is constructed by package *mgcv*:

$$\boldsymbol{\gamma}_z \sim \mathbf{MVN}(\mathbf{0}, \lambda_z^{-1}\mathbf{Q}_z^{-1}), \tag{12}$$

where $\mathbf{Q}_z$ is the precision matrix for the $z$th block of random effects $\boldsymbol{\gamma}_z$ and $\lambda_k$ is the estimated wiggliness (inverse-variance) parameter for that block. Computing the joint log-likelihood is efficient given that $\mathbf{Q}_{sem}$, $\mathbf{Q}_{dsem}$, $\mathbf{Q}_{spatial}$, and $\mathbf{Q}_z$ are all sparse. We use R-package *TMB* to apply the Laplace method to approximate the marginal likelihood, and optimize the log-marginal likelihood using gradients computed using automatic differentiation (Kristensen et al. 2016).

**Case-study: Habitat associations in data-integrated species distribution model**

To illustrate the wide range of models that can be expressed with *tinyVAST*, we introduce an "integrated species distribution model (SDM)" that combines samples from two surveys: a towed camera that measures densities of flatfishes (*Pleuronectiformes spp.*), rockfishes (*Sebastes* spp.), corals, and sponges, and a separate bottom trawl that measures densities of flatfishes and rockfishes. Individually, historical studies looking at species associations have often found positive correlations between rockfishes and corals and sponges in both trawl and camera data (Yoklavich et al. 2000, Laman et al. 2015). However, these two gears typically sample different types of substrates since trawling is limited to soft bottom substrates and at different spatial scales, since the area swept by an individual trawl haul is typically at least two orders of magnitude larger than a camera transect. So, it is unclear whether the strength of the positive impact on rockfish density is similar between the two gears. The opposite question is appropriate for flatfishes, which prefer soft-bottom sediments, are more likely to be effectively trawled, but are less likely to have sponges and corals present. We use high-level classification for fishes to avoid additional pre-processing associated with errors in species identification in the towed camera. Previous research has typically fitted an SDM to coral and sponge densities separately, predicted densities at the location of bottom trawls, and used those predictions as fixed covariates to explain flatfish and rockfish associations with biogenic habitat (Laman et al. 2018). By contrast, we seek to answer the following questions:

1. Are the associations between fishes and biogenic habitat estimated to be the same in towed camera and bottom trawl sampling programs?
2. Do the towed camera and bottom trawl measure similar variation in density for fishes, or does one of the other tend to measure more or less fluctuation in density between high and low-density areas due to density-dependent catchability for the bottom trawl (Kotwicki et al. 2014)?

The first set of data used in these analyses were from biennial (or triennial) bottom trawl surveys conducted from 1990 to 2019 in the Gulf of Alaska and the Aleutian Islands (von Szalay and Raring 2020, von Szalay and Raring 2018). The surveys are designed to assess the abundance of commercially important groundfishes and invertebrates in each ecosystem using a stratified random sampling design covering an area from Dixon Entrance in Southeast Alaska to Stalemate Bank in the Aleutian Islands at depths from 35 to 1000 m in the GOA and 35 to 500 m in the Aleutian Islands. Detailed sampling designs and protocols can be found in Stauffer (2004). In this analysis, catch and effort data from 14,877 research survey bottom trawl tows were used. The second set of data came from towed stereo-camera surveys conducted from 2012-2017 in the central and western Gulf of Alaska and Aleutian Islands from Yakutat, Alaska to Stalemate Bank (including Bowers Ridge and Bank). These data were collected using standardized sampling protocols of 15 minutes of on-bottom time where all fishes and benthic invertebrates were identified and counted to the lowest possible taxonomic level. The Aleutian Islands data were from 216 transects chosen using a stratified random survey design covering depths of ~15 to 850 meters (Rooper et al. 2016). The Gulf of Alaska data were from 227 transects chosen using a random or haphazard design (Sigler et al. 2022). The total number of each species of fish and invertebrate observed as well as the area of the seafloor viewed were calculated for each transect using the methods outlined in Rooper et al. (2016). However, in the analysis, both trawl survey and camera data used high-level classification of fish to avoid potentially erroneous species identifications in the camera data.

To answer these questions, we specify the following structural equations:

$$\log(\mu_S) = \alpha_S + \omega_S$$
$$\log(\mu_C) = \alpha_C + \omega_C$$
$$\log(\mu_F) = \alpha_F + b_1\omega_C + b_2\omega_S + \omega_F$$
$$\log(\mu_R) = \alpha_R + b_3\omega_C + b_4\omega_S + \omega_R \quad (13)$$
$$\log(\mu_{F2}) = \alpha_{F,t} + d_1\omega_C + d_2\omega_S + a_1\omega_F + \omega_{F2}$$
$$\log(\mu_{R2}) = \alpha_{R,t} + d_3\omega_C + d_4\omega_S + a_2\omega_R + \omega_{R2},$$

where $\mu_S$, $\mu_C$, $\mu_F$, and $\mu_R$ are the modeled numerical density of sponges, corals, flatfishes, and rockfishes in the towed camera, $\mu_{F2}$ and $\mu_{R2}$ are the modeled numerical densities of flatfishes and rockfishes in the bottom trawl, $\alpha_\blacksquare$ represents the intercept for each variable (which varies among years $t$ for the bottom trawl surveys but is constant for the drop-camera surveys), $\omega_\blacksquare$ represents spatial variation in log-density for a given variable, $b_1$ through $b_4$ represent habitat associations measured in the towed camera survey, $d_1$ through $d_4$ represent habitat associations between towed camera habitat and bottom trawl fish measurements, and $a_1$ and $a_2$ represent a nonlinear (density-dependent) detectability for the bottom trawl relative to towed camera measurement for fishes.

We then define a set of eight hypothesized relationships that arise as nested submodels. For example:

1. *Biogenic habitat as covariates*: We explore a model where sponges and corals are not linked to towed camera densities for fishes ($b_1 = b_2 = b_3 = b_4 = 0$). Instead, the sponges and corals are estimated as having a linear impact on bottom trawl estimates of fish densities (estimating $d_1$ through $d_4$), and other parameters are eliminated ($a_1 = a_2 = 0$). This then corresponds to a joint SDM that imputes the biogenic habitat densities and simultaneously estimates their use as covariates for predicting bottom trawl data.
2. *No link from drop camera and bottom trawl*: Alternatively, we explore a model where there is no link connecting drop camera and bottom trawl data ($d_1 = d_2 = d_3 = d_4 = a_1 = a_2 = 0$), while estimating habitat associations within the drop-camera data set. This measures the fine-scale association between habitat and fishes in the towed camera survey.
3. *Linked and spatially varying catchability*: Finally, we explore a model that assumes *a priori* that bottom trawl and drop camera samples are both measuring the same underlying variation in densities (i.e., $a_1 = a_2 = 1$), while also estimating residual variation in the detectability ratio (i.e., spatial variation in $\omega_{F2}$ and $\omega_{R2}$).

Other models are similarly formed via restrictions among parameters, which can be expressed using the "arrow notation" for interactions among variables.

Results suggest that the "Linked and spatially varying catchability" model is the most parsimonious of our hypothesized models (Fig. 1), although the saturated model and another candidate also receive some support ($\Delta AIC < 3$). This parsimonious model estimates a highly significant impact of sponges and rockfishes ($b_4 = 0.59$, $SE = 0.10$; Fig. 2), i.e., where a 10% increase in sponge densities is associated with a nearly 6% increase in rockfish density, and also estimates a significant and negative impact of corals on flatfishes ($b_1 = -0.26$, $SE = 0.05$). Model selection also indicates that it is parsimonious to specify that spatial variation in fish density in the towed camera is proportional to spatial variation observed using the bottom trawl ($a_1 = a_2 = 1$). Finally, the model estimates an "indirect" impact of corals and sponges on bottom-trawl fish densities via their estimated impact on towed-camera densities ($b_1$ through $b_4$) and the proportional relationship between towed camera and bottom trawl fishes ($a_1 = a_2 = 1$) but also a "direct" impact of biogenic habitat on bottom trawl fishes ($d_1$ through $d_4$). These "direct" impacts generally have a smaller magnitude than the indirect impacts, suggesting that the biogenic habitat has similar impacts for fish densities in both sampling programs. Finally, the model estimates spatial variation in the rockfish and flatfish densities calibrated to the bottom trawl (i.e., R2 and F2), and uses the drop-camera to estimate variable densities in Bower's Bank where the bottom trawl survey does not occur (Fig. 3). The bottom trawl is used to index density across space (to define essential fish habitat) and time (as input for stock assessments), so the estimated bottom-trawl densities in Bower's Bank represent useful information for fisheries managers.

**Discussion**

In this paper, we introduce the R-package *tinyVAST* to provide an expressive (general and flexible) interface for multivariate spatio-temporal models. This software includes functionality for a wide range of time-series and spatial models, including vector autoregressive, dynamic factor, empirical orthogonal function, and joint species distribution models (see Table 1). We intend to distribute this package via CRAN, and envision it as replacement for the package *VAST* (Thorson 2019) that is available via GitHub and is widely used in fisheries and marine sciences. It is "tiny" compared with *VAST* in that it uses Gaussian Markov random fields to represent all spatio-temporal variation, and thereby achieves a more condensed code base and more familiar user-interface. It has a similar interface to the package *sdmTMB* (Anderson et al. 2022), but includes additional options for multivariate analysis that are not available using that package. It also includes package vignettes that demonstrate the user-interface for a wide range of models (Table 5), and these vignettes compare results with alternative software including *sdmTMB* (Anderson et al. 2022), *VAST* (Thorson 2019), *mgcv* (Wood 2017), and *dsem* (Thorson et al. In press). We hope that this simple yet expressive interface will expand access to multivariate spatio-temporal models among applied ecologists.

**Acknowledgments**

Thanks to researchers from the AFSC Groundfish Assessment Program for collection of the historical bottom trawl survey data, and Rachel Wilborn, Kresimir Williams, Rick Towler and Darin Jones for their efforts in collecting and analyzing the underwater camera survey data. We also thank Arnaud Grüss, Mallarie Yaeger, and Nikolai Morokhovich for helpful comments on an earlier draft.

**Works cited**

Anderson, S. C., Ward, E. J., English, P. A. and Barnett, L. A. K. 2022. sdmTMB: an R package for fast, flexible, and user-friendly generalized linear mixed effects models with spatial and spatiotemporal random fields. - BioRxiv: 2022.03.24.485545.

Brooks, M. E., Kristensen, K., van Benthem, K. J., Magnusson, A., Berg, C. W., Nielsen, A., Skaug, H. J., Maechler, M. and Bolker, B. M. 2017. glmmTMB balances speed and flexibility among packages for zero-inflated generalized linear mixed modeling. - R J. 9: 378–400.

Charsley, A. R., Grüss, A., Thorson, J. T., Rudd, M. B., Crow, S. K., David, B., Williams, E. K. and Hoyle, S. D. 2023. Catchment-scale stream network spatio-temporal models, applied to the freshwater stages of a diadromous fish species, longfin eel (Anguilla dieffenbachii). - Fish. Res. 259: 106583.

Csardi, G. and Nepusz, T. 2006. The igraph software package for complex network research. - InterJournal Complex Syst. 1695: 1–9.

Fox, J., Nie, Z. and Byrnes, J. 2020. sem: Structural equation models. R package version 3.1-11.

Hoef, J. M. V., Peterson, E. and Theobald, D. 2006. Spatial statistical models that use flow and stream distance. - Environ. Ecol. Stat. 13: 449–464.

Ives, A. R., Dennis, B., Cottingham, K. L. and Carpenter, S. R. 2003. Estimating community stability and ecological interactions from time-series data. - Ecol. Monogr. 73: 301–330.

Kotwicki, S., Ianelli, J. N. and Punt, A. E. 2014. Correcting density-dependent effects in abundance estimates from bottom-trawl surveys. - ICES J. Mar. Sci. 71: 1107–1116.

Kristensen, K., Nielsen, A., Berg, C. W., Skaug, H. and Bell, B. M. 2016. TMB: Automatic differentiation and Laplace approximation. - J. Stat. Softw. 70: 1–21.

Laman, E. A., Kotwicki, S. and Rooper, C. N. 2015. Correlating environmental and biogenic factors with abundance and distribution of Pacific ocean perch (Sebastes alutus) in the Aleutian Islands, Alaska. - Fish. Bull. 113: 270–289.

Laman, E. A., Rooper, C. N., Turner, K., Rooney, S., Cooper, D. W. and Zimmermann, M. 2018. Using species distribution models to describe essential fish habitat in Alaska. - Can. J. Fish. Aquat. Sci. 75: 1230–1255.

Latimer, A. M., Banerjee, S., Sang Jr, H., Mosher, E. S. and Silander Jr, J. A. 2009. Hierarchical models facilitate spatial analysis of large data sets: a case study on invasive plant species in the northeastern United States. - Ecol. Lett. 12: 144–154.

Lindgren, F. 2023. fmesher: Triangle Meshes and Related Geometry Tools.

Lindgren, F., Rue, H. and Lindström, J. 2011. An explicit link between Gaussian fields and Gaussian Markov random fields: the stochastic partial differential equation approach. - J. R. Stat. Soc. Ser. B Stat. Methodol. 73: 423–498.

Ovaskainen, O., Tikhonov, G., Dunson, D., Grøtan, V., Engen, S., Sæther, B.-E. and Abrego, N. 2017. How are species interactions structured in species-rich communities? A new method for analysing time-series data. - Proc R Soc B 284: 20170768.

Pearl, J. 2009. Causal inference in statistics: An overview. - Stat. Surv. 3: 96–146.

R Core Team 2023. R: A Language and Environment for Statistical Computing. - R Foundation for Statistical Computing.

Thorson, J. T. 2019. Guidance for decisions using the Vector Autoregressive Spatio-Temporal (VAST) package in stock, ecosystem, habitat and climate assessments. - Fish. Res. 210: 143–161.

Thorson, J. T. and Kristensen, K. 2016. Implementing a generic method for bias correction in statistical models using random effects, with spatial and population dynamics examples. - Fish. Res. 175: 66–74.

Thorson, J. T. and Kristensen, K. 2024. Spatio-Temporal Models for Ecologists. - CRC Press.

Thorson, J. T., Andrews, A. G., Essington, T. and Large, S. In review. Dynamic structural equation models synthesize ecosystem dynamics constrained by ecological mechanisms.

van der Meer, L., Abad, L., Gilardi, A. and Lovelace, R. 2023. sfnetworks: Tidy Geospatial Networks.

Ver Hoef, J. M., Hanks, E. M. and Hooten, M. B. 2018. On the relationship between conditional (CAR) and simultaneous (SAR) autoregressive models. - Spat. Stat. 25: 68–85.

Wikle, C. K. and Cressie, N. 1999. A dimension-reduced approach to space-time Kalman filtering. - Biometrika 86: 815–829.

Wood, S. N. 2017. Generalized additive models: an introduction with R. - CRC press.

Wootton, J. T. and Emmerson, M. 2005. Measurement of interaction strength in nature. - Annu. Rev. Ecol. Evol. Syst.: 419–444.

Wright, S. 1934. The method of path coefficients. - Ann. Math. Stat. 5: 161–215.

Yoklavich, M. M., Greene, H. G., Cailliet, G. M., Sullivan, D. E., Lea, R. N. and Love, M. S. 2000. Habitat associations of deep-water rockfishes in a submarine canyon: an example of a natural refuge. - Fish. Bull. 98: 625–625.

Zuur, A. F., Ieno, E. N., Walker, N., Saveliev, A. A. and Smith, G. M. 2009. Mixed effects models and extensions in ecology with R. - Springer.

Table 1: List of model types and software functionality (rows) and R-packages (columns), indicating what functionality is supported by each R-package

| Model type | sem | mgcv | dsem | sdmTMB | VAST | tinyVAST |
|---|---|---|---|---|---|---|
| Available on CRAN | X | X | X | X | | X |
| *No spatial or temporal correlations* | | | | | | |
| Instrumental variables | X | | | | | X |
| *Time-series correlations* | | | | | | |
| Vector autoregressive model | | | X | | | X |
| Dynamic factor analysis | | | X | | | X |
| *Spatio-temporal correlations* | | | | | | |
| Spatial vector autoregressive model | | | | | X | X |
| Empirical orthogonal function analysis | | | | | X | X |
| Spatial factor analysis | | | | | X | X |
| Univariate species distribution model | | X | | X | X | X |
| Multivariate species distribution model | | X | | | X | X |
| Smoothing spline response for covariates | | X | | X | | X |
| Multiple distribution / link functions for different variables | | | | | X | X |

Table 2: Examples of common time-series models and their specification using "arrow-and-lag" notation. We envision data are available for two variables $X$ and $Y$ for vector autoregressive models (VAR) and dynamic factor analysis (DFA), or one variable $X$ for other models, and DFA and autoregressive integrated moving average (ARIMA) models involve estimating a latent variable $F$. All models are assumed to estimate an exogenous and independent variance for each variable by default. However, this default can be overridden by explicitly specifying a two-headed arrow where, e.g., "$X \leftrightarrow X$, 0, NA, 0" specifies that exogenous variance for variable $X$ is fixed at 0 *a priori*. Note that the DFA, ARIMA(1,1,0) and ARIMA(0,0,1) specifications are all reduced rank (arising from variable $X$ specified to have zero variance).

| Model and specification using "arrow-and-lag" notation | Explanation for each line |
|---|---|
| *Vector autoregressive model (VAR)* | |
| $X \rightarrow X$, 1, b_xx | First-order autoregression for $X$ (density dependence) |
| $X \rightarrow Y$, 1, b_xy | Lagged impact of $X_{t-1}$ on $Y_t$ |
| $Y \rightarrow X$, 1, b_yx | Lagged impact of $Y_{t-1}$ on $X_t$ |
| $Y \rightarrow Y$, 1, b_yy | First-order autoregression for $Y$ (density dependence) |
| *Dynamic factor analysis (DFA)* | |
| $F \rightarrow F$, 1, NA, 1 | Random walk for factor $F$ |
| $F \rightarrow X$, 0, b_fx | Loading of $X$ on factor $F$ |
| $F \rightarrow Y$, 0, b_fy | Loading of $Y$ on factor $F$ |
| $F \leftrightarrow F$, 0, NA, 1 | Unit variance for factor $F$ |
| $X \leftrightarrow X$, 0, NA, 0 | Turn off process error for $X$ |
| $Y \leftrightarrow Y$, 0, NA, 0 | Turn off process error for $Y$ |
| *ARIMA(1,1,0)* | |
| $F \rightarrow F$, 1, rho | AR1 for factor $F$ |
| $X \rightarrow X$, 1, NA, 1 | Random walk for time-series $X$ (integrated component) |
| $F \rightarrow X$, 0, NA, 1 | Fixed unit loadings for factor F on time-series $X$ |
| $X \leftrightarrow X$, 0, NA, 0 | Turn off additional process error for time-series $X$ |
| *ARIMA(0,0,1)* | |
| $F \rightarrow X$, 0, NA, 1 | Unit loadings for factor $F_t$ on $X_t$ in time $t$ |
| $F \rightarrow X$, 1, rho | Estimated loadings for factor $F_{t-1}$ on $X_t$ in time $t$ |
| $X \leftrightarrow X$, 0, NA, 0 | Turn off additional process error for time-series $X$ |

Table 3 – List of all user-defined arguments passed to function `fit` (the core function in `tinyVAST`) including different user options when specifying these.

| Argument name | Object passed | What it does |
|---|---|---|
| formula | Formula defining response, exogenous predictors, and offset variables | Defines regression-style model including nonlinear transformations |
| data | Data frame containing response, predictor, and offset variables | Contains all sampling data |
| sem | Options:<br>1. Character scring parsed by `make_sem_ram(.)` internally<br>2. Output from `make_sem_ram(.)` run manually | Defines interactions among variables that are constant over time. By default, *tinyVAST* uses "arrow notation" parsed by `make_sem_ram(.)` |
| dsem | Options:<br>1. Character string parsed by `make_dsem_ram(.)`<br>2. Output from `make_dsem_ram(.)` run manually<br>3. Output from `make_eof_ram(.)` run manually<br>4. Manually constructed "reticular action model" representing associations among variables over time | Defines interactions among variables over time, including lagged and simultaneous effects. By default, *tinyVAST* uses "arrow-and-lag notation" parsed by `make_dsem_ram(.)`. However, other models can instead be specified using alternative constructors, e.g., empirical orthogonal functions (EOF) |
| family | A standard R family object or a list with names matching elements of `data` corresponding to user-specified error distributions, where each element specifies a family and link function for those distributions. | Allows model to be specified to fit multiple data types, or a single data type but with different dispersion parameters associated with different samples (e.g., different variance for each modeled variable) |
| space_columns | A character vector where `data[,space_columns]` indicates the spatial coordinates for each sample | Matched against spatial coordinates or domain listed in `spatial_graph` |
| spatial_graph | S3-object, where options include:<br>1. Topology used by SPDE method to define spatial correlations in 2D spatial domain, constructed using *fmesher* package (Lindgren 2023)<br>2. Topology used by Simultaneous Autoregressive (SAR) method to | Defines spatial domain (extent and resolution) and resulting basis functions describing spatial correlations, where space-variable interactions arise as separable processes from `spatial_graph` and `sem`, and space-time-variable |
| | define spatial correlations for areal data (or independence for a multivariatiate time-series analysis), constructed using *igraph* package (Csardi and Nepusz 2006)<br>3. Topology used by autoregressive stream-network model for network data, constructed using *sfnetworks* package (van der Meer et al. 2023)<br>4. Missing, defining a time-series model (i.e., all samples are associated with a single site) | interactions arise as separable processes from `spatial_graph` and `dsem` |
| variable_column | A string where `data[,variable_column]` indicates the variable for each sample | Matched against variables listed in `dsem` and `sem` |
| variables | Character vector that defines the set of model variables, included as a column of `data` and indicated by `variable_column` | Defines the set of modeled variables (e.g., indicating species in a multispecies model). |
| time_column | A string where `data[,time_column]` indicates the time for each sample | Matched against lags specified in `dsem` |
| times | Integer vector that defines the set of model times, included as a column of `data` and indicated by `time_column` | Defines the temporal domain |
| control | Tagged list defining advanced features, constructed by `tinyVASTcontrol(.)` | Allows detailed control over model specification and optimization, while avoiding extra clutter for causal users. |
| delta_options | List with elements delta_formula, delta_sem, and delta_dsem (following same format as formula, sem, and dsem) | Allowing users to specify options when specifying a delta model via argument `family` |

Table 4: List of functions exported by *tinyVAST*, including their name and purpose

| Function name | Purpose |
| --- | --- |
| tinyVAST | Fit a *tinyVAST* model |
| make_sem_ram | Construct a reticular action model (RAM) using "arrow notation" to construct the precision matrix for a structural equation model |
| make_dsem_ram | Construct a RAM using "arrow-and-lag notation" to construct the precision matrix for a dynamic structural equation model |
| make_eof_ram | Construct a RAM for an empirical orthogonal function (EOF) analysis |
| summary | Summarize output for different model components |
| print | User-friendly output from a fitted model |
| logLik | The marginal log-likelihood, e.g., to enable function AIC(.) |
| predict | Predictions in response or link-scale, or for link-scale for individual components |
| residuals | Compute deviance or response residuals from a fitted model |
| integrate_output | Compute plug-in or epsilon bias-corrected estimator (Thorson and Kristensen 2016) for a quantity calculated via Monte Carlo integration across a user-specified set of sampling locations (and associated covariates) for a single time. This can then compute an area-weighted abundance index and measure distribution shifts, range expansion, and spatial overlap among variables. |

Table 5 – Vignettes included in package *tinyVAST*, including what model structures are specified and whether results are compared with alternative packages.

| Vignette | What it demonstrates |
| --- | --- |
| Dynamic structural equation models | Specifying a dynamic structural equation models (DSEM), and a comparison with package *dsem* (Thorson et al. In press) |
| Comparison with mgcv | Specifying a generalized additive models (GAMs), and a comparison with package *mgcv* (Wood 2017) |
| Empirical orthogonal function analysis | Specifying an empirical orthogonal function (EOF) model, which cannot be represented using "arrow-and-lag" notation and instead involves function `make_eof_ram(.)` |
| Simultaneous autoregressive process | Specifying a simultaneous autoregressive (SAR) model, and a comparison with a multivariate time-series model |
| Spatial modeling | Specifying a spatial model in two-dimensional coordinates, and a comparison with package *mgcv* |
| Spatial factor analysis | Specifying a spatial factor analysis (SFA) model that estimates covariance among multiple variables |
| Stream network models | Specifying a spatial model on a stream network |
| Vector autoregressive spatio-temporal models | Specifying a univariate spatio-temporal model, including a comparison with *VAST* (Thorson 2019) and *sdmTMB* (Anderson et al. 2022), and a bivariate vector-autoregressive model |

Fig. 1 – Visualizing the hypothesized relationship among six variables (Towed camera: R=rockfish; F=flatfish; C=corals; S=sponges;  Bottom trawl: R2=rockfish; F2=flatfish) using `tinyVAST::summary(.)` to summarize path coefficients, and functions `graph_from_data_frame(.)` and `plot.igraph(.)` in package *igraph* for plotting, where each arrows correspond to a specified path coefficient (**P**, see Eq. 6).  Estimated path coefficients list the parameter name, while path coefficients that are fixed a priori are shown without a name. We also show the marginal AIC relative to the most parsimonious model (bottom of each panel) and number of fixed effects (top of each panel), and all models include intercepts for each sampled combination of variable and year (56 parameters), two dispersion parameters per variable ($2 \times 6 = 12$ parameters), one variance per species (6 parameters), and an estimated decorrelation rate (1 parameter) that are not shown here.

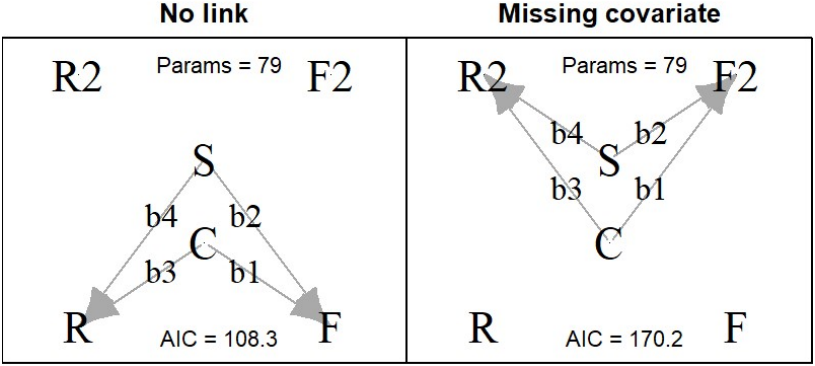
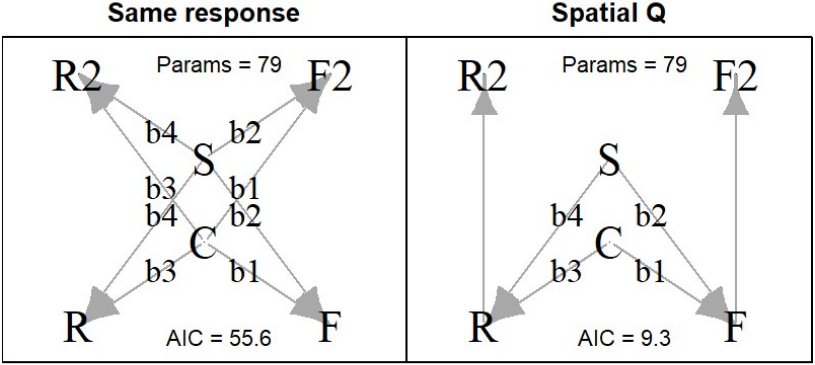
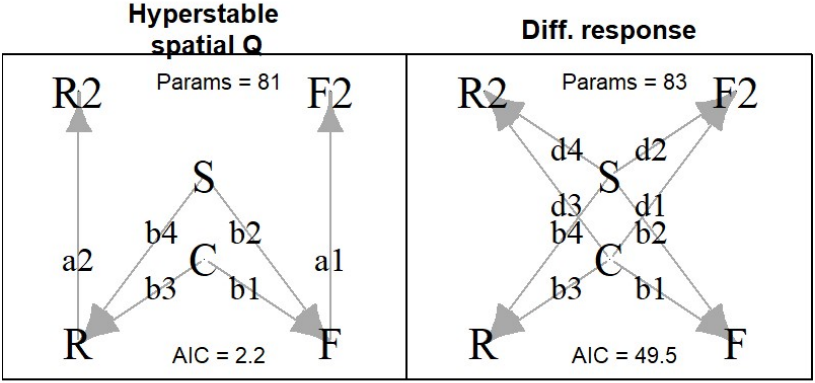
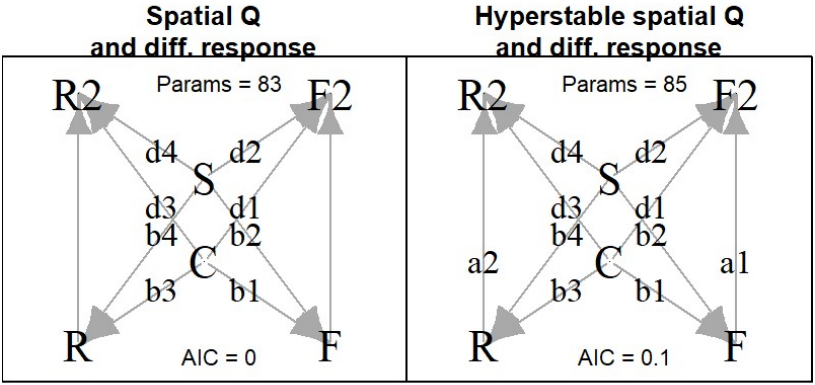

Fig. 2 – Estimated path coefficients for the most parsimonious model (bottom-left of Fig. 1), showing both maximum likelihood estimate and standard error (in parentheses), where standard error NA corresponds to parameters that are fixed a priori.

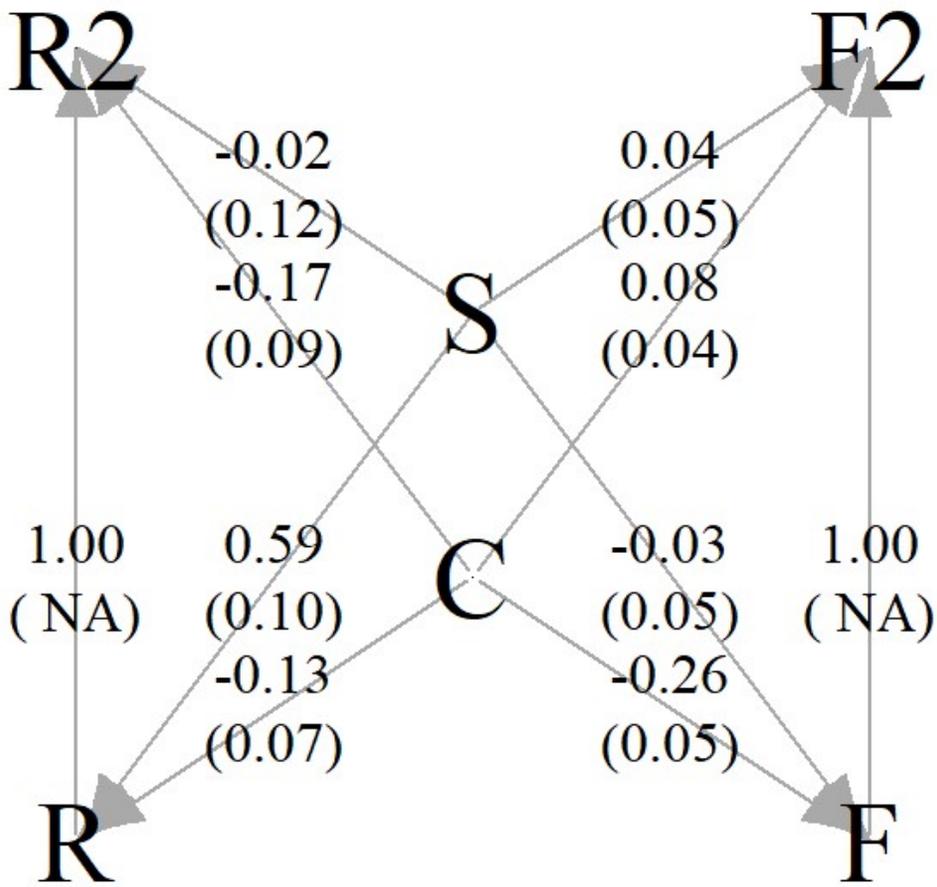

Fig. 3 – Estimated log-densities (e.g., "Coral" in top-left panel) and the standard error in log-density (e.g., "SE[ Coral ]") for each of six modeled variables (Coral, Sponge, Rock = rockfishes, and Flat = flatfishes in the drop camera, and Flat_trawl and Rock_trawl in the bottom trawl) showing the location of sampling data for that variable (green dots) on top of estimated standard errors. Here we specifically show densities in Bower's Bank (a relatively shallow area north of the central Aleutian Islands that is not sampled by the bottom trawl survey) to show that drop camera samples can predicted densities from the bottom trawl.

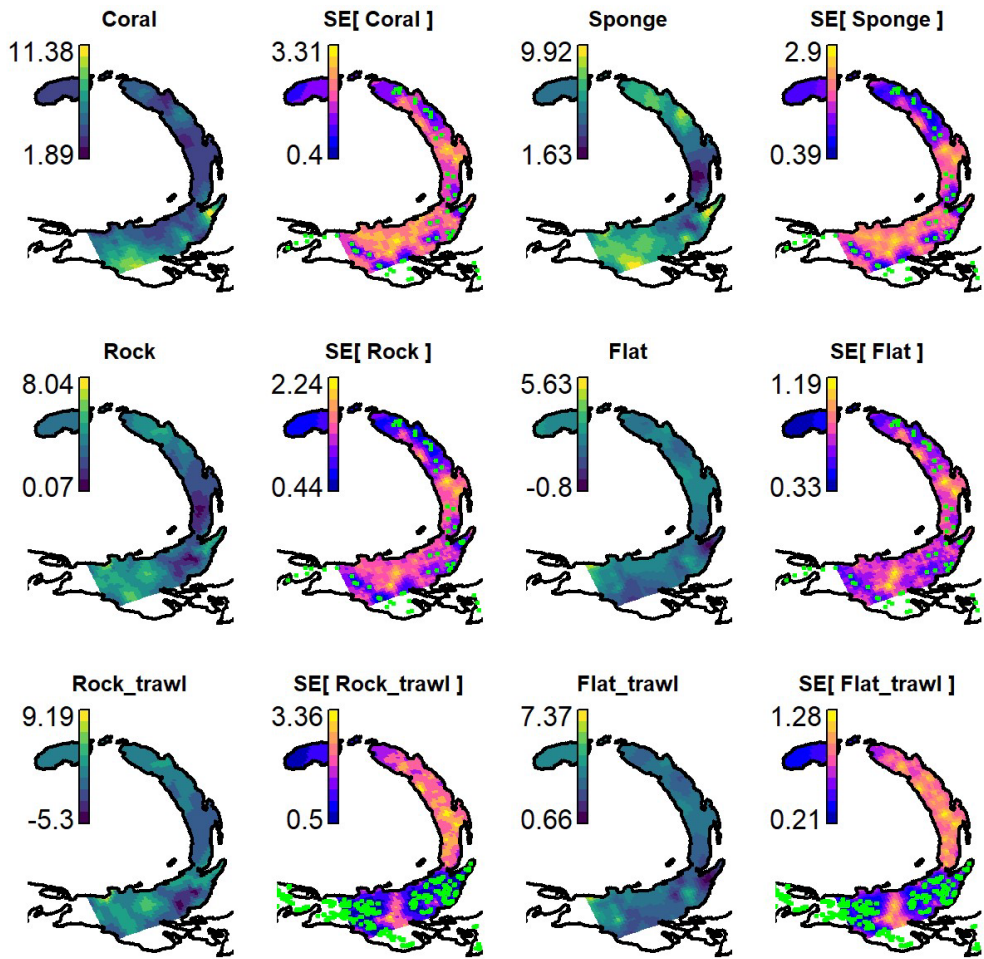